\documentclass{kluwer}	  % Specifies the document style.

\usepackage[dvips]{graphicx}

\begin{document}
\begin{article}
\begin{opening}
\title{Inner disk oscillations}
\author{Tomaso \surname{Belloni}}
\runningauthor{Tomaso Belloni}
\runningtitle{Inner Disk Oscillations}
\institute{Osservatorio Astronomico di Brera, Merate, Italy}
\date{October 26, 2000}

\begin{abstract}
The RXTE observations of GRS 1915+105 have given a new impulse
to the study of spectral and timing properties of the X-ray
emission of black hole candidates. At variance with any other
known source, GRS~1915+105 shows dramatic changes on time scales
as short as a tenth of a second. These changes are associated
to marked spectral changes which have been interpreted as
changes of the observable inner radius of the accretion disk.
I review the existing results and discuss the current evidence
for such disk oscillations. Independently of the precise
theoretical models, the detailed study of these fast variations
can provide an extremely valuable insight on the accretion
processes onto black holes. Making use of these results, I compare
the properties of GRS~1915+105 with those of other black hole
candidates.
\end{abstract}

\end{opening}

\section{Introduction: black hole candidates}

In the recent years, a classification scheme for the X-ray emission
of black hole candidates has emerged. From the timing and spectral
properties, four separate states (plus the quiescent state for
transient systems) have been identified, probably in dependence of
the accretion rate level (see van der Klis 1995). Unfortunately, all known
persistent sources, Cyg X-1, GX~339-4, LMC~X-1 and LMC~X-3, show
state transitions very rarely, making the accumulation of extensive
datasets on different states difficult. Transient systems are more
promising, but their transient nature limits the number of states
and state transitions that can be observed from one system, and the
comparison between different systems is always problematic.

The spectral properties of black hole candidates are usually
characterized in terms of a double-component model. The first
component is thermal and (when present) contributes mostly to
the flux below 10 keV. It is commonly interpreted as emission
from an optically thick accretion disk (Mitsuda et al. 1984).
The second component
is much harder and extends above 10 keV. In its simple form,
it can be approximated as a power law, although more sophisticated
models are often required (see e.g. Frontera et al. 2000).
An attractive feature of the optically thick disk model, the so-called
{\it disk-blackbody}, is that one of its parameters is the
value of the inner radius of the accretion disk. Therefore,
the application of
this model can in principle yield the measurement of this fundamental
parameter, although a precise value depends on the distance and
the inclination of the system. This model became rather popular
when a number of black hole systems showed a rather constant
inner disk radius, around 20-30 km, despite large changes in
accretion rate (see Tanaka \& Lewin 1995). Although this
model is simplified and leads to an underestimate of the
values of the radii, it can detect large variations in the inner
radius of the disk (see Merloni, Fabian \& Ross 2000).

\section{GRS~1915+105: inner disk oscillations}

The galactic Microquasar GRS~1915+105 is known to exhibit
dramatic variability in the soft (1-20 keV) energy band
(see Greiner, Morgan \& Remillard 1996; Belloni et al. 2000) 
and since its appearance in 1992
it has remained active up to the time of writing.
This variability is accompanied
by strong spectral changes (Greiner, Morgan \& Remillard 1996; Belloni
et al. 1997a,b; Muno, Morgan \& Remillard 1999), providing an ideal system to
study spectral transitions in a black-hole candidates: the
source is always observable and significant spectral variability
is present on very short time scales. At the same time, a
complex behavior is observed in the power density spectra of
the source (see Rao, this volume).
%---------------------------------------------------------------------
%\begin{figure}[H] % figuur 1
%\tabcapfont
%\centerline{%
%\begin{tabular}{c@{\hspace{6pc}}c}
%\includegraphics[width=1.0in,height=1.9in]{grs_fig1.ps} &
%\includegraphics[width=1.0in,height=1.7in]{grs_fig2.ps} \\
%\end{tabular}}
%\caption[]{Left: Count rate, inner disk temperature and radius as a function
	%of time of the 1997 June 18 observation of GRS~1915+105. 
	Right: corresponding
	%relation between disk re-fill time and inner disk radius. (from
	%Belloni et al. 1997b).}
%\end{figure}
%-----------------------------------------------------------------------
%---------------------------------------------------------------------
\begin{figure}
\centerline{
\includegraphics[width=18pc]{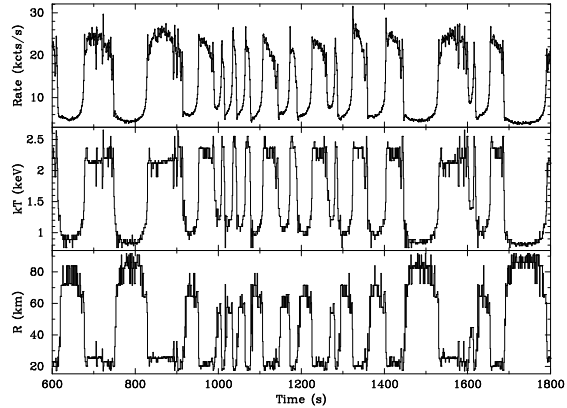}}
\caption{Count rate, inner disk temperature and radius as a function
        of time of the 1997 Jun 18th of GRS~1915+105 (from
        Belloni et al. 1997b).}
\end{figure}
%---------------------------------------------------------------------
%---------------------------------------------------------------------
\begin{figure}
\centerline{
\includegraphics[width=18pc]{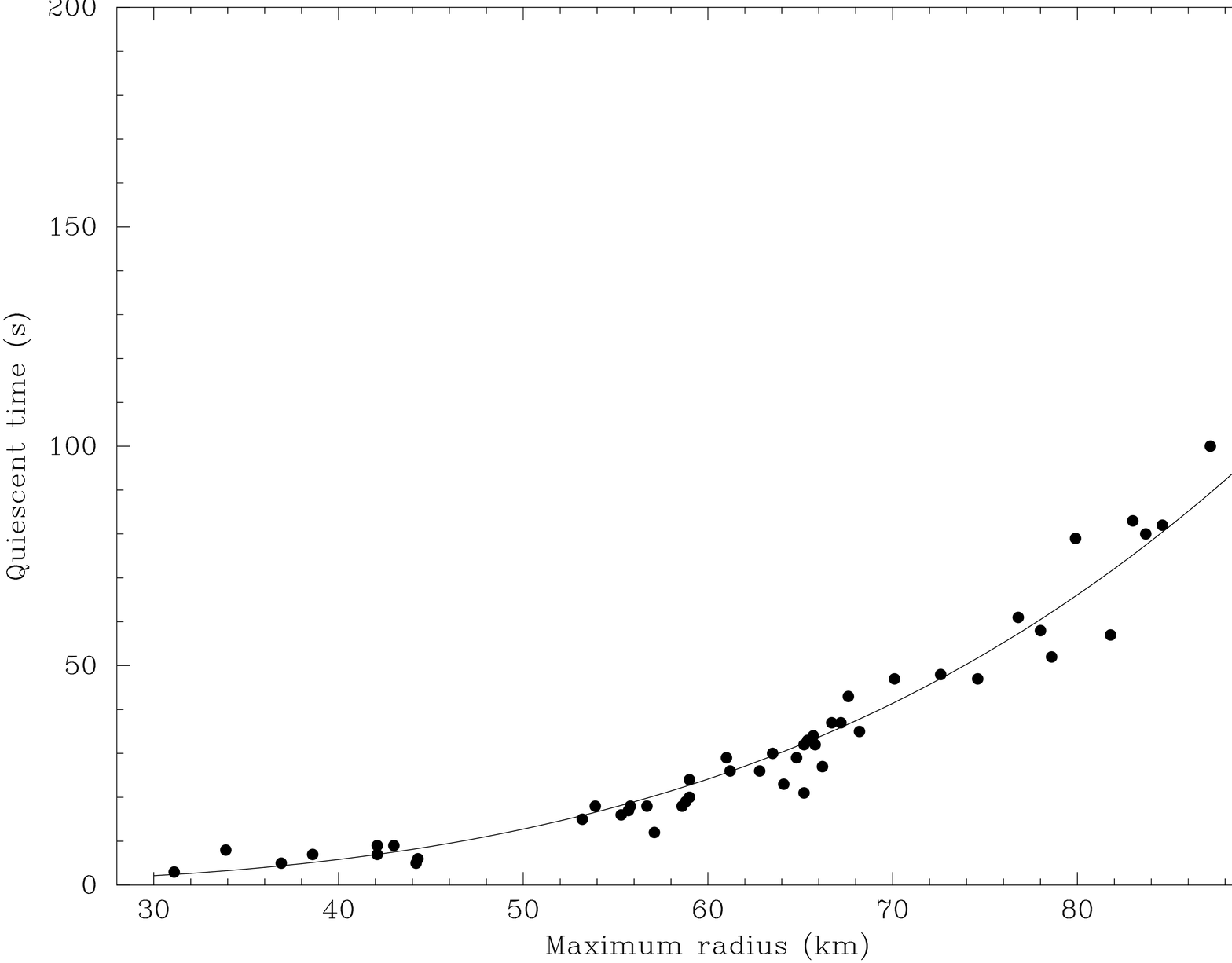}}
\caption{Relation between disk re-fill time and inner disk radius for
	the same observation shown in Figure 1 (from
	Belloni et al. 1997b).}
\end{figure}
%---------------------------------------------------------------------

Belloni et al. (1997a,b) proposed a model for the observed
spectral variations, based on fits to RXTE/PCA data with the approximated
spectral shape described above. They interpret the observations
as the onset of a thermal-viscous instability, during which the
innermost part of the accretion disk becomes unobservable
and is slowly refilled from the outer parts. This interpretation
is based on the measurement of a variable inner disk radius
from energy spectra (see Figure 1) and is supported by the
observation of the expected dependence between size of the
unstable region and refill time (Figure 2). The
observed radius variations are too large to be attributed to
spurious effects due to the approximate form of the model.
This is what I will refer to as `inner disk oscillations'.
The modeling by these authors was rather qualitative (with the
exception of the estimate of the refill time scale), but other
authors have developed more accurate models for this process
(Szuszkiewicz \& Miller 1998, Nayakshin, Rappaport \& Melia 2000,
Janiuk et al. 2000).

\section{Spectral states of GRS~1915+105}

Belloni et al. (2000) analyzed a large number of RXTE/PCA observations
of GRS~1915+105 and obtained a subdivision of the observed
phenomenology into 12 separate classes. From this classification,
they identified three basic states, the alternation of which cause
all of the observed variability. The three classes, called A, B and
C, are shown schematically in Figure 3, which represents a color-color
diagram (both colors increase with increasing hardness of the spectrum).

These states correspond to the three states already identified by
Markwardt, Swank \& Taam (1999). Class B correspond to the ``normal''
state of a black hole candidate at high accretion rate (the very high
state), with an optically thick accretion disk extending down to the
last stable orbit and a steep power-law component. State C
corresponds to the instability periods: the accretion disk stops at
a larger radius than in state B, while the power law component is
harder. State A is a new state, not recognized earlier. It corresponds
to an accretion disk like in state B, but with a lower temperature,
and therefore lower local accretion rate. The relatively fast
transition time between A and B is consistent with being the
viscous time scale at the innermost stable orbit around the black hole.
The observed variability
consists of transitions between these three states: all possible
transitions between two of the states are observed, with the exception
of C to B transitions, of which no example has been found.
%---------------------------------------------------------------------
\begin{figure}
\centerline{
{\rotatebox{270}{\includegraphics[width=18pc]{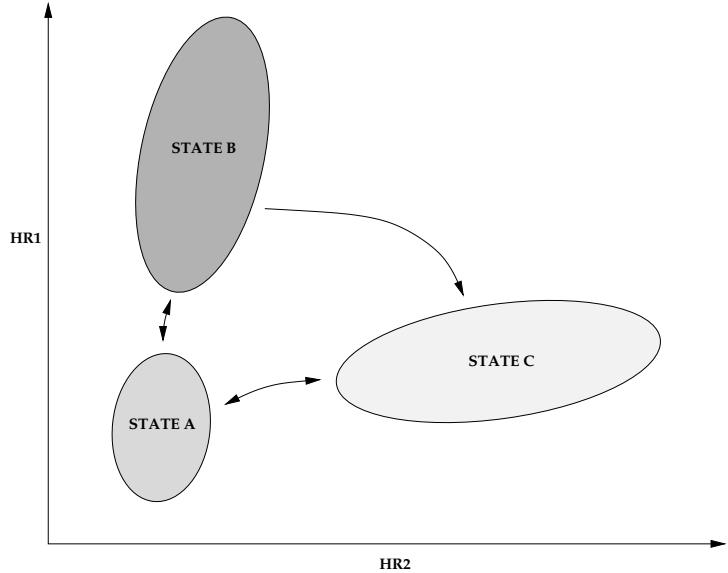}}}}
\caption{Schematic color-color diagram showing the basic A/B/C states
        and their observed transitions (from Belloni et al. 2000).}
\end{figure}
%---------------------------------------------------------------------

There are intervals of time, even as long as a month, when the source
is found only in state C. These are the `plateau' intervals observed
in the radio band (see Fender 1999). These are consistent with being
long instability intervals, when a large missing inner disk is
slowly refilled at a lower external accretion rate (Belloni et al. 2000).
On the other hand, there are very fast variations as well. During
some observations, variability of a factor of five in 0.1 seconds
has been observed, corresponding to fast A--B/B--A transitions (Belloni
et al. 2000).

\section{Timing properties}

It is important to connect what is observed in the time domain and
the state transitions described above. As a first step, I consider the
three types of QPOs observed in GRS~1915+105:

\begin{itemize}

\item 1-10 Hz QPO: (see Rao, this volume). This QPO is only observed
	during state C. Its central frequency varies systematically
	with time, count rate and hardness. Since state C is the only
	state when systematic changes in the inner disk radius are
	observed, it is natural to associate this oscillation to the
	inner radius, although often the missing part of the disk is
	so large that no disk component is observed directly, indicating
	that the QPO must be related to the power law component.

\item 67 Hz QPO: This QPO appears only during state B. Its high
	and rather constant frequency
	and its association with a state with small (and possibly constant)
	inner disk radius also point to a connection to the inner disk
	radius. It is interesting to note that this QPO is not observed
	in all state-B intervals, indicating that there must be another
	parameter involved in its production.

\item Low-$\nu$ QPO: These oscillations, in the range 10-100 seconds
	(see Morgan, Remillard \& Greiner 1997) {\it are} the regular
	transitions between states that are often observed. The
	instability model provides an interpretation for the variability,
	but does not cast light on why the light curves are so regular.

\end{itemize}

There is another important point about the timing properties of the source.
Looking at the light curves (see Belloni et al. 2000), one notices not only
that their time structure is very complex, but also it often repeats
in an almost undistinguishable way, so that all observations can be
grouped in twelve classes. As mentioned above, there is no clear explanation
for why the structure of the light curves is the observed one.
More than this, sometimes even the finest
structures of the light curves repeat at a distance of years (see
Figure 4). The basic question that need to be answered are:
why do the light curves have those complex and repeatable shapes, and
why do the light curves have {\it only} those complex shapes, with
only a few possibilities to choose from? Answering these questions
could provide crucial information about the accretion phenomenon.

%---------------------------------------------------------------------
\begin{figure}
\centerline{
\includegraphics[width=18pc]{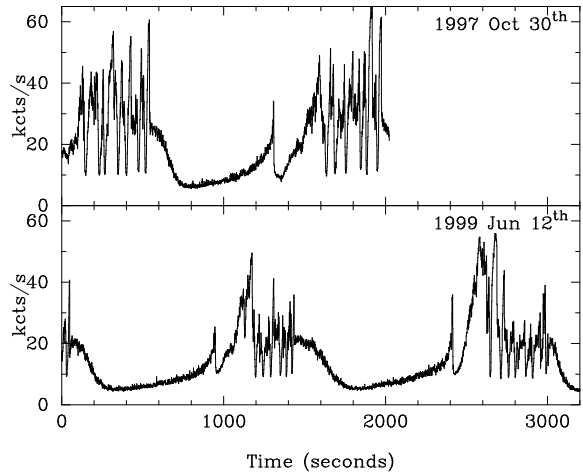}}
\caption{Two PCA light curves of GRS~1915+105 separated by almost two
	years. The similarity between the two panels, even in some
	small details, is evident.}
\end{figure}
%---------------------------------------------------------------------

\section{A unique source?}

Another important question is: are there other sources that show
the same phenomenology seen here? To this date, the answer is:
the C-state events (the instability) are observed only in
GRS~1915+105, but A--B transitions have been observed in other
systems, namely 4U~1630-47 and GRO~J1655-40 (Trudolyubov,
Borozdin \& Priedhorsky
2000, see Figure 5). Possibly, the dips and the so-called
flip-flops seen earlier in GX~339-4 (Miyamoto et al. 1991) are
also of the same nature. It seems clear that this type of
fast (and often very regular) temperature oscillations are
more common among bright black hole candidates, although their
origin is still unknown.
The question of why the C-state instability
is observed only in this source is still basically unanswered.
Possibly, this type of instability appears only at very high
levels of accretion rate, which would then be reached only by
GRS 1915+105. The observation of the same phenomenon in another
source would greatly help to understand this issue.

\section{Relation to the canonical black hole states}

An important point is the comparison between the A/B/C states of
GRS 1915+105 and the four canonical black hole states mentioned
above. It is tempting to associate state C with the low/intermediate
state (hard spectrum flat-top noise and QPO in the power spectrum),
state B with the very high state (strong disk component, weaker
noise level), and the
A state with the high state (cooler disk component, low noise level).
However, as mentioned above, the instability related to the C state
is not observed in other sources. It is more likely that the
similarities between the properties of GRS~1915+105 and the
canonical black hole states are not indicative of them being the
same, but rather of them looking the same. In other words, the
onset of the instability in GRS~1915+105 influences the accretion
structure in a way that makes it mimic the properties of the
canonical states observed in ``normal'' sources.

%---------------------------------------------------------------------
\begin{figure}
\centerline{
\includegraphics[width=18pc]{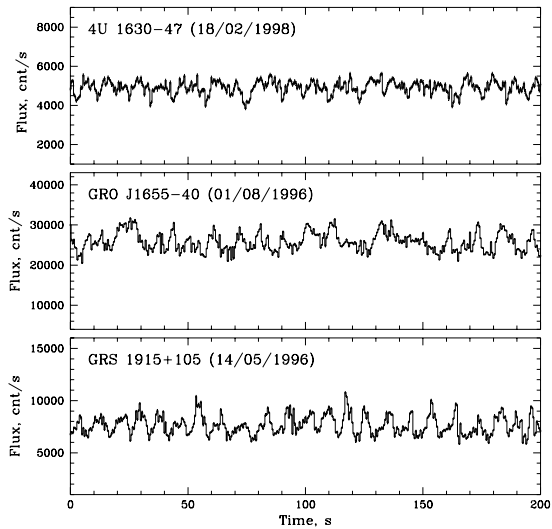}}
\caption{Comparison between selected RXTE/PCA light curves of 4U 1630-47
(upper panel, GRO J1655-40 (middle panel) and GRS 1915+105 (lower panel).
From Trudolyubov, Borozdin \& Priedhorsky (2000).}
\end{figure}
%---------------------------------------------------------------------

\section{Inner disk oscillations and radio jet ejection}

The X-ray phenomenology described above has been positively linked to the
variability observed in the radio band, and major events have been
associated to the emergence of superluminal radio jets. This topic is
treated by other authors in this volume. Here it is important to stress that
the analysis of simultaneous X-ray/radio observations has shown that 
radio flares seem to be associated only to state C events, and therefore to 
the inner disk oscillations, while observations containing only states A and
B do not correspond to significant radio detections of the source 
(see Klein-Wolt et al., this volume). However, the presence of this
instability has not yet been observed in the other galactic sources showing
jet ejection in the radio.

\section{Conclusions}

GRS~1915+105 is the only source up to now which can provide a large
number of spectral transitions, enabling us to study different states
of one source without having to wait for years. The most important of
these transitions do not involve the canonical
states of black hole candidates, but are associated to an instability
of the innermost region of an optically thick accretion disk, which
causes its inner parts to be evacuated and refilled on the local
viscous time scale. Other variations
observed in this source can be linked to those observed
in other sources, strengthening the connection between this unique
system and more conventional ones. On the X-ray side, it is now
important to examine the spectral/timing behavior of GRS 1915+105
in more detail (see Migliari, Vignarca \& Belloni, this volume),
in order to provide a more complete phenomenological picture for
further theoretical work. Modeling of the X-ray
properties of GRS~1915+105 is at the moment the most promising
way to make significant progress in the understanding of accretion
onto black holes.

% The endnotes section will be placed here.

\theendnotes

\end{article}

\begin{thebibliography}{}

\bibitem{}
Belloni, T., et al., 1997a, ApJ, 479, L145

\bibitem{}
Belloni, T., et al., 1997a, ApJ, 488, L109

\bibitem{}
Belloni, T., et al., 2000, A\&A, 355, 271

\bibitem{}
Fender, R.P., et al., 1999, MNRAS, 304, 865

\bibitem{}
Frontera, F., et al., 2000, ApJ, in press (astro-ph/0009160)

\bibitem{}
Greiner, J., Morgan, E.H., Remillard, R.A., 1996, ApJ, 473, L107

\bibitem{}
Janiuk, A., et al., 2000, ApJ, in press (astro-ph/0008354)

\bibitem{}
Markwardt, C.B., Swank, J.H., Taam, R.E., 1999, ApJ, 513, L37

\bibitem{}
Morgan, E.H., Remillard, R.A., Greiner, J., 1997, ApJ, 482, 993

\bibitem{}
Merloni, A., Fabian, A.C., Ross, R.R., 2000, MNRAS, 313, 193

\bibitem{}
Muno, M.P., Morgan, E.H., Remillard, R.A., 1999, ApJ, 527, 321

\bibitem{}
Mitsuda, K., et al., 1984, PASJ, 36, 741

\bibitem{}
Miyamoto, S., et al., 1991, ApJ, 383, 784

\bibitem{}
Nayakshin, S., Rappaport, S., Melia, F., 2000, ApJ, 535, 798

\bibitem{}
Szuszkiewicz, E., Miller, J.C., 1998, MNRAS, 298, 888

\bibitem{}
Tanaka, Y., Lewin, W.H.G, 1995, in ``X-ray binaries'',
		  Cambridge Univ. Press, p126

\bibitem{}
Trudolyubov, S.P., Borozdin, K.N., Priedhorsky W.C., 2000,
	MNRAS, in press (astro-ph/9911345)

\bibitem{}
van der Klis, M., 1995, in ``X-ray binaries'', Cambridge Univ. Press, p252

\end{thebibliography}
\end{document}